# REFRACTIVE INDEX OF NANOSCALE THICKNESS FILMS MEASURED BY BREWSTER REFRACTOMETRY

\* E. A. Tikhonov, Senior SPIE member, Dr. phys.-math. sci., \* Lyamets A. K., senior engineer, \*\*Bespalova I. I., Ph. deg. phys.-math.sci.,\*\* Malyukin Yu. V., Dr. phys.-math. sci.
\* Institute of physics of the National Academy of Sci., Kiev, Ukraine
\*\* Institute of scintillation materials of National Academy of Sci., Kharkov, Ukraine
E-mail: etikh@live.ru

Abstract - It is shown that reflective laser refractomtery at Brewster angle can be useful for precision measurements of refractive indexes (RI) in the transparency band of various films of nanoscale thickness. The RI measurements of nanoscale porous film on the basis of gadolinium orthosilicate and quartz glass have been carried out as first experience. It is shown that surface light scattering in such films that is connected with clustering of nanoscale pores can decrease the accuracy of the RI measurements at Brewster angle. Estimated physical dependence RI stipulated by the film thickness reduction (3D-2D transition) in the range of (20-160)nm has not been not detected.

## 1. INTRODUCTION REMARKS

At all variety of the developed methods refractometry [1] they are virtually unusable for precision measurements of refractive index (RI) of nanometer thickness films due to the occurrences of the substrate. RI measurement of such objects can be implemented with the help of methods and devices for laser ellipsometry [2]. However, in compare to the larger number of measured parameters in similar process it is more practical to turn to the laser refractometry at Brewster angle (LRBA) [3]. Advantages of LRBA, in particular, are enclosed in the measurement of single parameter - Brewster angle and in the definite absence of restrictions on such features of films as bulk light scattering and absorption under conditions that the imaginary part of the complex RI remains at least one order smaller than the its real part [3].

In this work we turned to Brewster refractometery RI of porous nanoscale films with thicknesses in the range from 20 to 300nm manufactured by the sol-gel technology of precipitation from solution on the quartz substrate, followed by annealing to enable formation of nanoscale pores. The aim of this work was to measure possible changes of RI films, when their thickness will become less than the testing wavelength. Expected changes of RI with a thickness are possible due to the transition of samples from a 3-dimensional (3D) to 2-dimensional state (2D) [4] and because of affecting the pore concentration, homogeneity of its distribution and sizes with thickness in the used technology of manufacturing.

## 2. MOTIVATION OF GOALS, SELECTION OF SAMPLES AND MEASUREMENT TECHNIQUES

In Lorentz-Lorenz model for samples described as the mixture of several components refraction is determined by the sum of their individual refraction in accordance with known relation [4]:

$$\frac{n^2-1}{n^2+2}\left(\frac{m}{\rho}\right) = \frac{n_1^2-1}{n_1^2+2}\left(\frac{m_1}{\rho_1}\right) + \frac{n_2^2-1}{n_2^2+2}\left(\frac{m_2}{\rho_2}\right) + .. \quad (1)$$

where m, n, ρ mass, refractive index and specific weight of mixture components, taking up volumes $V_1+V_2+..=V$.

In the case of porous materials the considered components are quartz glass and air. To determine the dependence of the effective refractive index of the porous glass on the concentration of air we can take the approximation $n_2=1$. Then from the expression (1), we obtain the following approximate relation:

$$\left(\frac{n^2-1}{n^2+2}\right)V = \left(\frac{n_1^2-1}{n_1^2+2}\right)V_1 \quad (2)$$

This ratio allows to represent the dependence of RI porous glass as the function of the relative specific volume $\Delta=V_1/V$, occupied by the solid phase:

$$n(\Delta) = \sqrt{1+2\Delta\left(\frac{n_1^2-1}{n_1^2+2}\right)\bigg/1-\Delta\left(\frac{n_1^2-1}{n_1^2+2}\right)} \quad (3)$$

It is clear that with the growth of the total volume occupied by pores, RI of porous glass will tend to 1. The scattering light in the case the pores of nanometric sizes is considered the insignificant one by definition.

In order the calculations were to be consistent with refraction measurements at the Brewster angle, the level of light scattering with increasing thicknesses of experimental samples should not change significantly [3]. For a typical values RI n for solid silicate glasses n=1,5; 1,6; 1,8; 2,0 calculated dependence n=f($\Delta$ x) for porous analogues with the growth of the pore concentration demonstrates trend RI to 1 (Fig.1.)

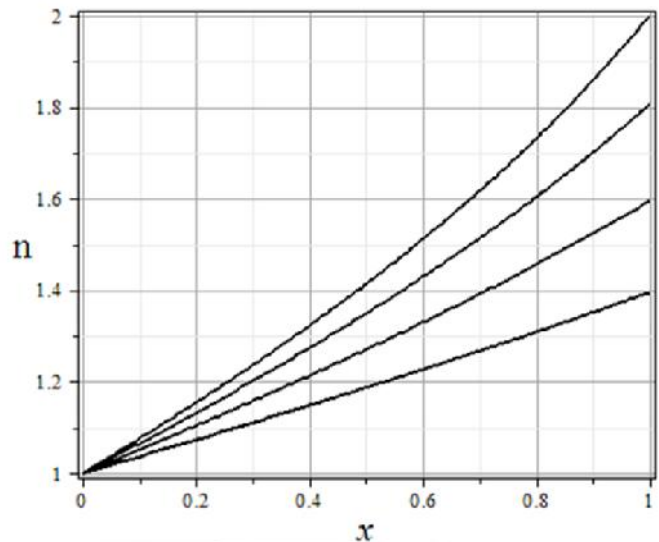

Fig.1. RI - pore concentration dependence of porous glass.
RI of solid material (rightmost) is taken equal 1,4; 1,6; 1,8; 2,0.

There are the available references about the phenomenon of transition of thin films 3D→ 2D state with the thickness decrease and the hints about possibility of registration of this transition by measuring RI. But itself nanoscale thickness is not enough physical reason for the film transition in 2-dimensional state. The most significant criteria the dimension of the state of the film is the change in the temperature dependence of the heat capacity [4]. So, for the nanoscale crystalline film with incommensurable periods of nuclear topography relative to the substrate the heat capacity - temperature relation in the 2D state varies according to a quadratic law, while for the samples of the similar film in 3-dimensional state the same relation is characterized by a cubic dependence.

To identify the state dimensionality of the film exist other criteria also: for example, the critical temperature $T_C$ relation for the phase transition of films is equal to 2D/3D=0,4 [4]. Similar changes of thermodynamic parameters qualitatively explained by the change of lattice coordination number, which determines the number of neighboring atoms of the crystal lattice of bulk and thin film samples. Changes in the intensity of the reflected light with the film thickness of krypton, is given in [4], also indicate the change in RI with film thickness that can be associated with changes in specific weight density. Perhaps just this factor that causes the different RI on a surface and inside volume indicated by the calculations and the experiment conducted in Rayleigh time to explain nonvanishing reflection at Brewster's angle for p-polarized radiation [5]. In this case, the use of models of the Lorentz-Lorenz for nanoscale films coming from 3D to 2D-dimensional condition is getting inadequate.

In our experiments we studied the behavior of RI nanoporous fused quartz and films of nanoscale thickness on the basis of gadolinium orthosilicate ($Ga_2SiO_5$-GSO) as function of thickness and porosity.

Porous fused glass represent a significant practical interest for applications in photonics and optics [6,7,8], including their application as matrices for fluorescent labels on rare earth elements [9]. Volume porous silica and porous films of GSO were prepared using sol-gel technique followed by annealing of the synthesized material at temperatures of 900-10000C. In the absence of annealing the material remained solid. Porous and solid film GSO of different thicknesses were obtained by sequential deposition from a solution on a quartz substrate the required number of layers (1,4,8,16,30) with a thickness of 20 nm.

Contrary to expectations, the optical homogeneity of the porous films were relatively low, which is manifested in the low reproduction of the RI measurement results. About distribution uniformity of the arising pores and defects across surface/volume in our films we can judge on the visually analyzed pattern of laser light scattering and by pictures registered with a scanning electron microscope under the high spatial resolution (see fig.2. below).

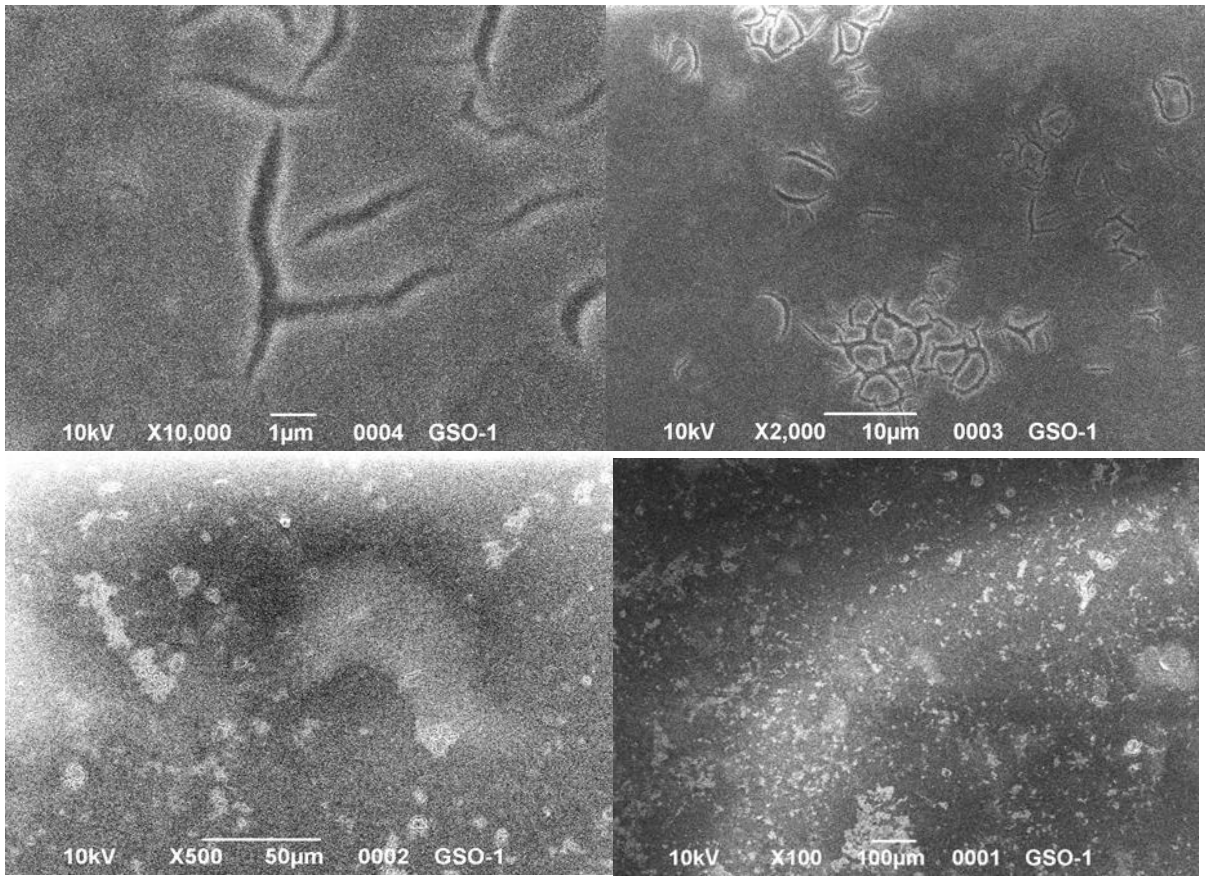

Fig.2. Film of gadolinium orthosilicate GSO-1 thickness 20nm registered with increased resolution from upper left to down right (in 100 times): dim region of uniform illumination correspond to areas with low scattering light, their size does not exceed 50mkm.

Provided photos demonstrate that even the most homogeneous samples GSO have only small homogeneous regions 50mkm. While laser beam used in our RI measurements had the cross-section 1 mm and therefore repeatedly overlapped the some homogeneous regions associated with the peculiarities of the separated pore distributions and their concentration gradients. This led to the averaged power of the reflected beam in the vicinity of Brewster angle and determined Brewster angle as the angular position of reflected power minimum. Scattering light, which in the case of the nanometric pore sizes would almost be absent, due to their inhomogeneous distribution and grouping into clusters larger 1 mkm takes place. As GSO film thickness increase with the number of deposited layers, the total thickness of the layer and the heterogeneity increases also, consequently, the level of light scattering increased, that was manifested in the reduction of the signal-to-noise relation due to summing the contributions of the mirror (coherent) and dispersed by the angle components of the reflected light.

According to the above mentioned formula (3) at the given pore concentration the RI film value doesn't depend from thickness. However, in the our experimental case of the film thickness increase have been followed the light scattering growth so that measured variations RI value may connected with next reasons:

1. The transverse size of the testing laser beam exceeded the dimensions of the homogeneous areas of the film 50 mk, that resulted in some averaging of RI value.

2. Comparable levels of the scattering light and the useful signal at the film thickness increase can affect the angular position of Brewster minimum which determines RI values.

For all measurements RI of nanoscale thickness films on transparent substrates by Brewster refractometry, it was also necessary to eliminate the contribution of power, mirrored from back face of the substrate by blackening or antireflection coating it or setting aperture for the reflected beam (if the film thickness is larger than the measuring beam diameter). In the absence of these precautions the angular dependence of the reflected power from the working and substrate back surface will be sum of useful and given false powers. In the case of materials with different RI of the film and substrate total dependence will form the minimum reflected power at an angle that is different from the real Brewster angle of the film [1]. Especially harmful for accurate measurements is the depolarized component the scattered light at the Brewster angle. In our method it was suppressed by the polarization analyzer that was tuned to pass only the p-component of the reflected beam.

### 3. THE MEASUREMENTEL RESULTS AND THEIR DISCUSSION

Let us refer to actual measurements RI of porous materials by laser Bruwster refractometry. In Fig.3. presents the angular dependence of the reflected power p-polarized radiation of a He-Ne laser (632.8 nm) of the bulk fused quartz with a porosity of 50%, determined by the method of Russian state standard of [10]. The sample had a relatively low level of Relaigh light scattering, because the wave front of specular reflected measuring beam was accompanied by very low intensive diffuse background of scattered light with a low degree of depolarization: this allowed us to determine Brewster angle with an accuracy appropriate to surfaces of optical quality that provides an average roughness about $\lambda/10$. The data of Fig.3. provide the expected reduction RI=1,259 relatively solid quartz (PP=1,456), that is consistent with the calculated dependence in Fig.1., pointing to the integral pore volume of 40%. Highest decrease of RI magnitude 1,08 for porous quartz was achieved in work /9/.

The decrease in RI due to the porosity is accompanied by a decrease of the level of Fresnel reflected power on the at all angles of incidence. As a result the useful signal in the vicinity of Brewster angle was below the noise level (at a given power several mW of the used laser). Therefore, to restore all the dependency including the region in the vicinity Brewster angle, the polynomial regression analysis was used from the soft OriginPro 8.0".

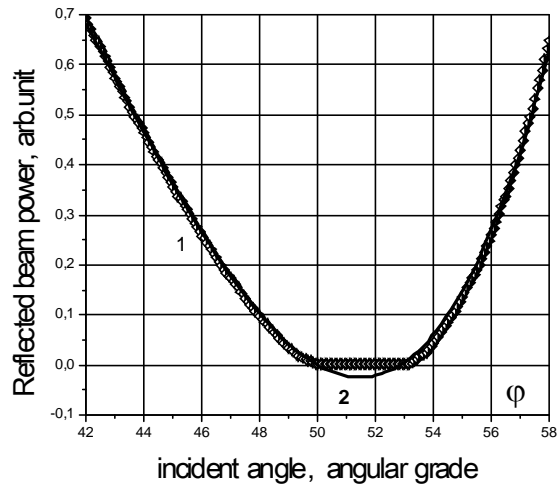

Fig.3. The light reflected power from the surface of the porous quartz at a wavelength of 632.8 nm (He-Ne laser in the p-configuration of linear polarization, curve 1). To determine Brewster angle is used polynomial regression at correlation coefficient 0.998, $\varphi_{br}=51,535^0$ and RI=1,259 (curve 2).

RI measurement of the nanoscale thickness porous films were conducted with the material on the basis of gadolinium orthosilicate. Increasing level of light scattering with the growing film thickness was restricted with values up to 160nm because an increase film thicknesses the ratio of specular reflected power to diffuse one was getting no sufficient to restore lost in the noise of a wide area in the vicinity of Brewster angle.

The results of measurements on RI on Brewster angle at the wavelength of $\lambda$=660nm diode laser is presented in Fig.4.

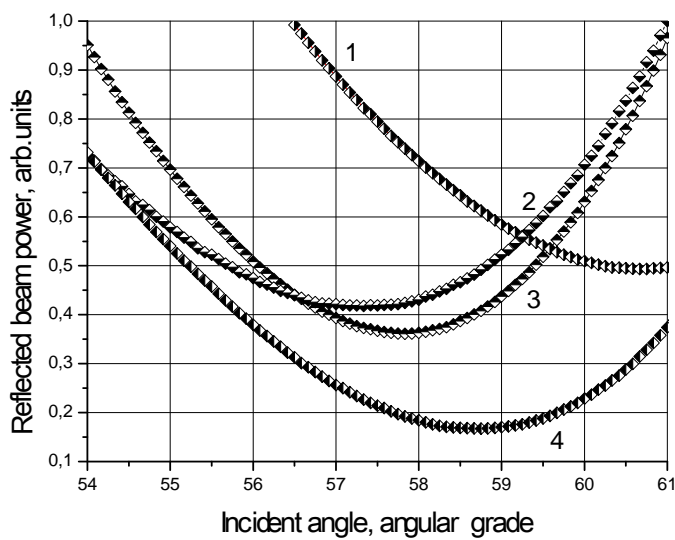

Fig.4. Determination of RI of the porous gadolinium orthosilicate films by Brewster angle refractometry : 1 – 80nm, RI=1,780 (solid), 2- 80nm, RI=1,559 (porous), 3-20nm, RI=1,590 (porous), 4-160nm, RI=1,648 (porous); laser wavelength 660±5 nm.

Results show that RI of all porous film samples are considerably less then RI=1,780 of solid gadolinium orthosilicate film. However, the values of the measured RI differ from estimated according to Lorentz-Lorenz model with homogeneous distribution of the equal sized nanopores. Moreover, averaging of inhomogeneity, which took place in the measurements with the beam, the cross section of which exceeded the heterogeneity, was not enough to even the averaged RI values for samples used in measurements. Averaging over the 3 samples gives RI=1.6 that forecasts the porosity of film samples equal 20% according to above presented calculation on Fig.1.

If to compare the dependences obtained with the porous fused quartz (Fig.3.), one can see that the reflected power level on Brewster angle for all samples of gadolinium orthosilicate (Fig.4.) due to strong light scattering is more significant due to the large optical inhomogeneities and higher RI of the solid material.

## 4. CONCLUSION

Presented results allow to make the following conclusions:

1. The possibility of measurement RI of nanoscale thicknesses films using the laser refractometric methods at Brewster angle is demostrated. The technique remains operational ability even for samples at relatively strong diffusion light scattering, however, under conserving the specular components of the incident beam of light in the reflected beam.

2. The analysis of the RI trend in porous films with thickness variation on the basis of the specific refraction Lorentz-Lorenz model gives a qualitative and quantitative description of the results.

3. Possible application of this technique for samples nanoscale thicknesses, with emphasis on the role of the transition layer material in the formation reflected beam power at Brewster's angle has been pointed.

4. The results of RI changes of the porous films within the boundaries of the measurement of their thicknesses (20-160)nm not gives grounds to conclude that the transition of the film in 2D-dimensional state takes place, indicating on search of that with the films of subnanoscale thickness.